\documentclass[preliminary,copyright,creativecommons]{eptcs}

\providecommand{\event}{TERMGRAPH 2016} 

\usepackage{amsmath}
\usepackage{macros,graphicx,pict2e,verbatim,nets}
\usepackage{alltt,multicol}

\newcommand     \ar             {{\sf ar}}
\newcommand     \inter          {\bowtie}
\newcommand     \myvec          {\mathaccent "017E}
\newcommand     \EQ             {\mathbin{=}}

\newcommand{\ito}{\Rightarrow}         
\newcommand{\Zero}{\texttt{Z}}
\newcommand{\Succ}{\texttt{S}}
\newcommand{\add}{\texttt{Add}}
\newcommand{\sym}[1]{\mathtt{#1}}
\newcommand     \Lra            {\Rightarrow}
\newcommand\nil{{\sf nil}}
\newcommand\rev{{\sf{rev}}}

\newcommand{\normalscale}{0.48}
\newcommand{\smallscale}{0.42}
\newcommand{\tinyscale}{0.40}

\newcommand{\Func}[1]{\mathsf{#1}}

\newcommand\ind[1] {\${#1}}

\newcommand{\NameSet}{\mathcal{N}}

\newcommand{\Empty}{-}

\newcommand{\delimStrS}{\mbox{``}}
\newcommand{\delimStrE}{\mbox{''}}
\newcommand{\Str}[1]{\delimStrS{}#1\delimStrE}

\begin{document}

\title{In-place Graph Rewriting with Interaction Nets}
\def\titlerunning{In-place Graph Rewriting with Interaction Nets}  
\author{Ian Mackie\institute{LIX, CNRS UMR 7161, \'Ecole Polytechnique, 91128 Palaiseau Cedex, France} \and Shinya Sato\institute{University Education Center, Ibaraki University, 2-1-1 Bunkyo, Mito, Ibaraki 310-8512, Japan}}
\def\authorrunning{I. Mackie \& S. Sato}
\maketitle

\begin{abstract}
An algorithm is in-place, or runs in-situ, when it does not need any
additional memory to execute beyond a small constant amount. There are
many algorithms that are efficient because of this feature, therefore
it is an important aspect of an algorithm.  In most programming
languages, it is not obvious when an algorithm can run in-place, and
moreover it is often not clear that the implementation respects that
idea. In this paper we study interaction nets as a formalism where we
can see directly, visually, that an algorithm is in-place, and
moreover the implementation will respect that it is in-place.  Not all
algorithms can run in-place however. We can nevertheless still use the
same language, but now we can annotate parts of the algorithm that can
run in-place. We suggest an annotation for rules, and give an algorithm
to find this automatically through analysis of the interaction rules.
\end{abstract}

\section{Introduction}

An algorithm runs in-place, or in-situ, if it needs a constant amount
of extra space to run.  For an algorithm to be in-place, the input is
usually overwritten, so mutable data-structures need to be supported
by the programming language. There are many well-known in-place
algorithms, in particular from the domain of sorting. One example is
bubble sort, that we can write in Java for instance:
\begin{center}\small
\begin{verbatim}
  static void bubble() {
    int t;
    for (int i = n-1; i >= 0; --i)
      for (int j = 1; j <= i; ++j)
        if (a[j-1] > a[j]) 
          { t = a[j-1]; a[j-1] = a[j]; a[j] = t; }
  }
\end{verbatim}
\end{center}

With some knowledge of what the above instructions do, and tracing a
few steps of the execution, we can soon realise that it runs in-place:
one additional memory location (\texttt{t}) is all that is needed to
sort the array \texttt{a} of integers.  In many programs however, it
is not obvious that an algorithm can run in-place, and moreover it is
often not clear that the underlying implementation respects that
idea. This issue becomes more pertinent when we examine different
programming paradigms and different programming styles, especially
when we have dynamic data-structures.

In Figure~\ref{fig:programs} we give four fragments of programs for
inserting an element into a sorted list (so part of the insertion sort
algorithm). These programs are written in Prolog, Haskell and Java,
with the latter written using two different programming
styles. Insertion can be written so that it runs in-place, but it is
not easy to see which of these examples are (or can be) in-place
unless we start to examine the compiler, the run-time system, and the
definition of functions like \texttt{cons} in the case of
Java. Declarative languages (functional and logic based in this
example) are designed to be referentially transparent, so the
data-structures are updated in a non-destructive way. Moreover, it is
not the programmer who decides how memory is allocated and organised
in these languages. On the other-hand, languages like C and Java (the
imperative fragment) the programmer does the memory allocation (and
de-allocation also in some languages) explicitly, and therefore has a
better idea of resource usage.  These examples illustrate some of the
difficulties in knowing if the program will run in-place or not.

\begin{figure}[htbp]
\begin{minipage}[t]{.45\textwidth}\small
  Program 1: Prolog
\begin{verbatim}
insert([Y|Ys], X, [Y|Zs]):- 
  Y < X, !, insert(Ys, X, Zs).
insert(Ys, X, [X|Ys]).
\end{verbatim}
\end{minipage}\hfill
\begin{minipage}[t]{.45\textwidth}\small
  Program 2: Haskell
\begin{verbatim}
insert e [] = [e]
insert e (x:xs) = if e < x then e:x:xs 
  else x:(insert e xs)
\end{verbatim}
\end{minipage}

\vspace{1cm}

\begin{minipage}[t]{.45\textwidth}\small
  
  Program 3: Java (functional style)
\begin{verbatim}
static List insert(int x, List l) {
  if (isEmpty(l) || x < l.head) 
    return cons(x, l);
  else
    return cons(l.head, insert(x, l.tail));
}
\end{verbatim}
\end{minipage}
\hfill
\begin{minipage}[t]{.45\textwidth}\small
  
  Program 4: Java (destructive style)
\begin{verbatim}
static List insert(int x, List l) {
  if (isEmpty(l) || x < l.head)
    return cons(x, l);
  else {
      l.tail = insert(x, l.tail);
      return l;
  }
}
\end{verbatim}
\end{minipage}
\caption{Example programs}\label{fig:programs}
\end{figure}

In-place algorithms are important because they can lead to more
efficient algorithms, and even change time complexity. If the
data-structure supports it, concatenation of two lists can be constant
time if done in-place, but linear if not.  Memory allocation is also
expensive, so minimising it also makes it more efficient. Although
less important in some ways, there are devices that have limited space
(embedded systems, hand-held devises, etc.), so limiting the space
usage if there is no run-time impact is always advantageous.

In this paper we use a formalism where we can see directly, in fact
visualise, that an algorithm is in-place, and moreover the
implementation respects that it is in-place. We use the graphical
rewriting system of interaction nets~\cite{LafontY:intn} as our
programming paradigm. This visual language has many similarities with
term rewriting systems in that they are user-defined systems. For this
reason, they can be considered as specification languages. However,
they are also a model of computation that requires all aspects of the
computation to be explained, including copying and garbage
collection. For this reason they are like an implementation model, or
low-level language. It is the mixture of these features that allows us
to see directly how the program can be implemented, and thus see how
the memory is allocated.

Not all algorithms can run in-place however. The formalism will still
be of use though, and we identify three different uses of the
information we can ascertain from interaction rules:

\begin{enumerate}
\item If the rewrite rules have a particular property then the
  algorithm is in-place (and will be implemented in-place).

\item If the rules can be applied in a given way, so a strategy is
  needed, then the algorithm can be implemented in-place. 

\item If neither of the above hold, then we can still make use of the
  formalism by re-using as much data as possible in the computation.
\end{enumerate}

For the final point, we can either ask the programmer to annotate the
rules, or develop an algorithm to do this automatically. In this
paper, we give examples to motivate the first and last points---more
details including the second point will be given in a longer version
of this paper.

The space usage of algorithms, as well as the time complexity, are
fundamental in algorithm design and analysis, and well documented in
many textbooks. There have also been a number of works that give a
bound on the space usage through type systems, for
example~\cite{HofmannM:typsb}, and \cite{HughesP99}. Our approach is
more syntactical, and uses properties of the underlying run-time
system.

\paragraph{Overview.}
The rest of this paper is structured as follows. In the next section
we recall the background, and give some examples to motivate the
ideas. We then give some case studies of examples that are in-place in
Section~\ref{sec:inplace}. In Section~\ref{sec:annotation} we
introduce an annotation for the rules which allows for node reuse. We
then go on in Section~\ref{sec:deriving} to show an algorithm to
annotate a rule automatically in the case of using a fixed-size node representation for nodes.
After a brief discussion on how we can
use this information in a compiler in Section~\ref{sec:discussion}, we
conclude in Section~\ref{sec:conc}.

\section{Background}

In the graphical rewriting system of interaction
nets~\cite{LafontY:intn}, we have a set $\Sigma$ of \emph{symbols},
which are names of the nodes. Each symbol has an arity $ar$ that
determines the number of \emph{auxiliary ports} that the node has. If
$ar(\alpha) = n$ for $\alpha \in \Sigma$, then $\alpha$ has $n+1$
\emph{ports:} $n$ auxiliary ports and a distinguished one called the
\emph{principal port} labelled with an arrow.  Nodes are drawn as
follows:
\begin{center}
\includegraphics[width=1.6cm,keepaspectratio,clip]{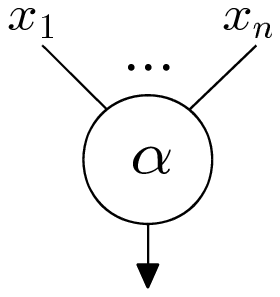}
\end{center}

A \emph{net} built on $\Sigma$ is an undirected graph with nodes as
the vertices.  The edges of the net connect nodes together at the
ports such that there is only one edge at every port.  A port which is
not connected is called a \emph{free port}.  Two nodes
$(\alpha,\beta)\in \Sigma\times\Sigma$ connected via their principal
ports form an \emph{active pair}, which is the interaction net
analogue of a redex.  A rule $((\alpha, \beta) \ito N)$ replaces the
pair $(\alpha, \beta)$ by the net $N$.  All the free ports are
preserved during reduction, and there is at most one rule for each
pair of nodes.  The following diagram illustrates the idea, where $N$
is any net built from $\Sigma$.
\begin{center}
\includegraphics[width=7.7cm,keepaspectratio,clip]{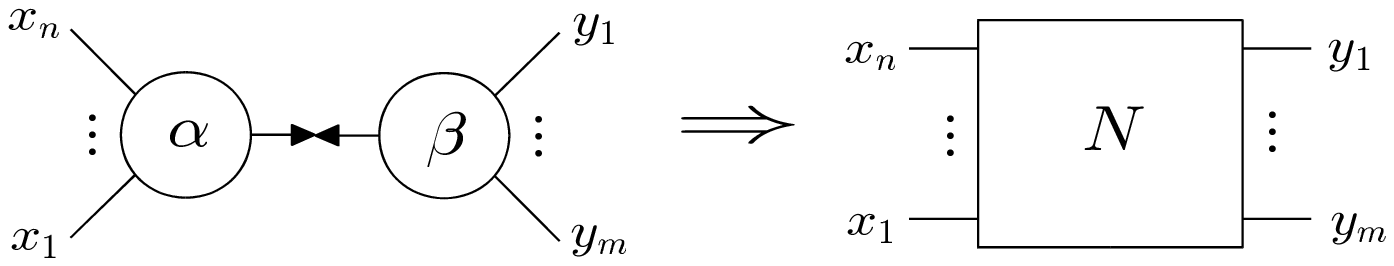}
\end{center}
We refer to the rule $((\alpha, \beta) \ito N)$ as
$\alpha\inter\beta$.  The most powerful property of this graph
rewriting system is that it is one-step confluent---all reduction
sequences are permutation equivalent. We use an extension of these
pure interaction nets: values can be stored in the nodes, and rules
can test these values. This is done is such a way as to preserve the
one-step confluence property. We use this extension in the insertion
sort example below.

It is possible to reason about the graphical representation of nets,
but it is convenient to have a textual calculus for compact
representation. There are several calculi in the literature, and here
we review one calculus~\cite{DBLP:journals/corr/HassanMS15}, which is
a refined version of \cite{MackieIC:calin}.

\begin{description}
\item[Agents:] Let $\Sigma$ be a set of symbols, ranged over by
  $\alpha,\beta,\ldots$, each with a given \emph{arity} $\ar : \Sigma
  \to \NAT$. An occurrence of a symbol is called an \emph{agent}, and
  the arity is the number of auxiliary ports.
  
\item[Names:] Let $\NameSet$ be a set of names, ranged over by $x,y,z$, etc.
  $N$ and $\Sigma$ are assumed disjoint. Names correspond to wires in the graph system.
  
\item[Terms:] A term is built on $\Sigma$ and $\NameSet$ by the
  grammar: $t\; \mathrm{::=}\; x \mid \alpha(t_1,\ldots,t_n) \mid \ind{t}$, \,
  where $x\in \NameSet$, $\alpha \in \Sigma$, $\ar(\alpha) = n$ and
  $t_1,\ldots,t_n,t$ are terms, with the restriction that each name can
  appear at most twice.  If $n = 0$, then we omit the parentheses.
  If a name occurs twice in a term, we say that it is \emph{bound},
  otherwise it is \emph{free}.  We write $s,t,u$ to range over terms,
  and $\vec{s}, \vec{t}, \vec{u}$ to range over sequences of terms.  A
  term of the form $\alpha(t_1,\ldots,t_n)$ can be seen as a tree with
  the principal port of $\alpha$ at the root, and the terms
  $t_1,\ldots,t_n$ are the subtrees connected to the auxiliary ports
  of $\alpha$. The term $\ind{t}$ represents an indirection node which
  is created by reduction, and is not normally part of an initial
  term. Intuitively, $\ind{t}$ corresponds to a variable bounded with $t$ (or a state such that an environment captures $t$).

\item[Equations:] If $t$, $u$ are terms, then the unordered pair
  $t\EQ u$ is an \emph{equation}. $\Theta$ will be
  used to range over sequences of equations. 

\item[Rules:] Rules are pairs of terms written: $\alpha(x_1, \ldots,
  x_n) \EQ \beta(y_1, \ldots, y_m) \ito \Theta$, where $(\alpha,\beta)
  \in \Sigma \times \Sigma$ is the active pair, and $\Theta$ is the
  right-hand side of the rule. We will abbreviate in the following the
  left- and right-hand sides of the rule by LHS and RHS
  respectively. All names occur exactly twice in a rule, and there is
  at most one rule for each pair of agents.

\end{description}

\section{In-place algorithms: case studies}\label{sec:inplace}

Here we give some example interaction net systems that demonstrate the
ideas we have discussed previously.  The first one is unary numbers
with addition.  We represent the following term rewriting system:
\verb|add(Z,y)=y, add(S(x),y)=add(x,S(y))| as a system of nets with
nodes $\Zero$, $\Succ$, $\add$, and two rewrite rules:
\begin{center}
\includegraphics[scale=\normalscale,keepaspectratio,clip]{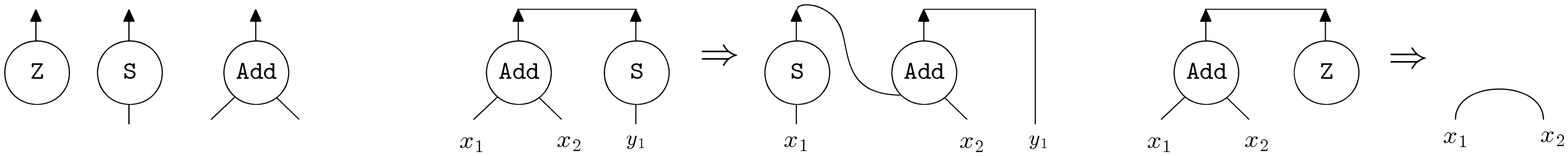}
\end{center}
The following is an example of \verb|add(S(Z),Z)|:
\begin{center}
\includegraphics[scale=\normalscale,keepaspectratio,clip]{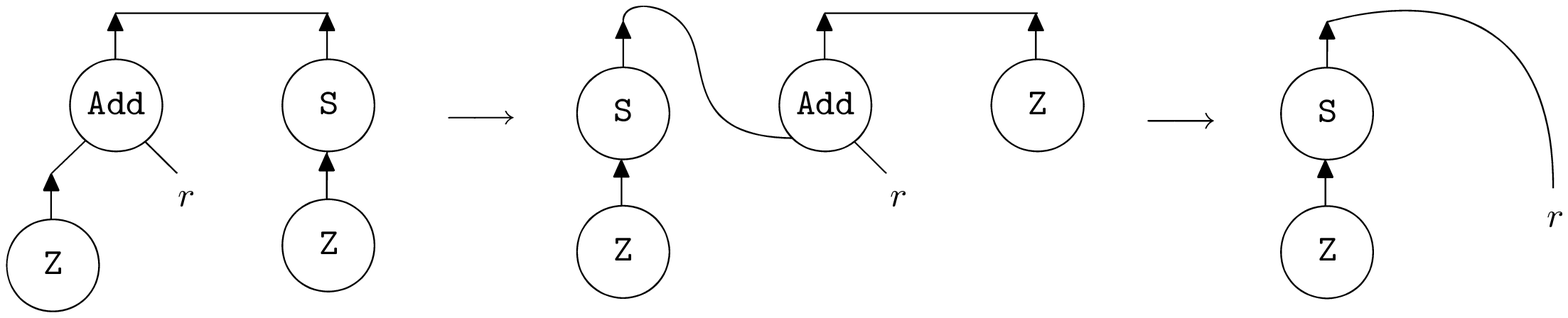}
\end{center}

Next, we introduce an example for the Ackermann function defined by:
\begin{verbatim}
 ack 0 n = n+1,  ack m 0 = ack (m-1) 1,  ack m n = ack (m-1) (ack m (n-1)).
\end{verbatim}
We can build the interaction net system on the unary natural numbers
that corresponds to the term rewriting system as follows:
\begin{center}
\includegraphics[scale=\normalscale,keepaspectratio,clip]{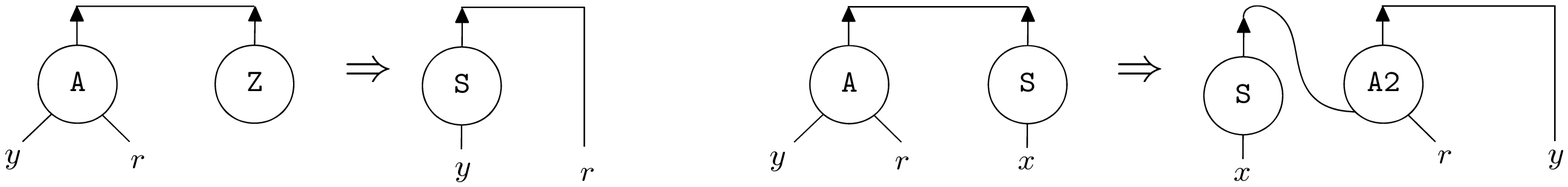}

\smallskip
\includegraphics[scale=\normalscale,keepaspectratio,clip]{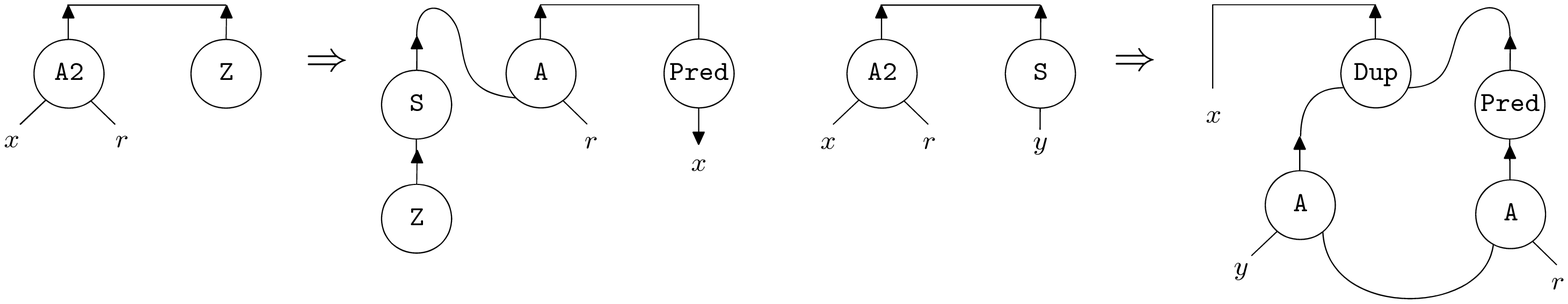}
\end{center}
where the node $\sym{Dup}$ duplicates $\sym{S}$ and $\sym{Z}$ nodes,
and the node $\sym{Pred}$ erases the $\sym{S}$ node:
\begin{center}
\includegraphics[scale=\smallscale,keepaspectratio,clip]{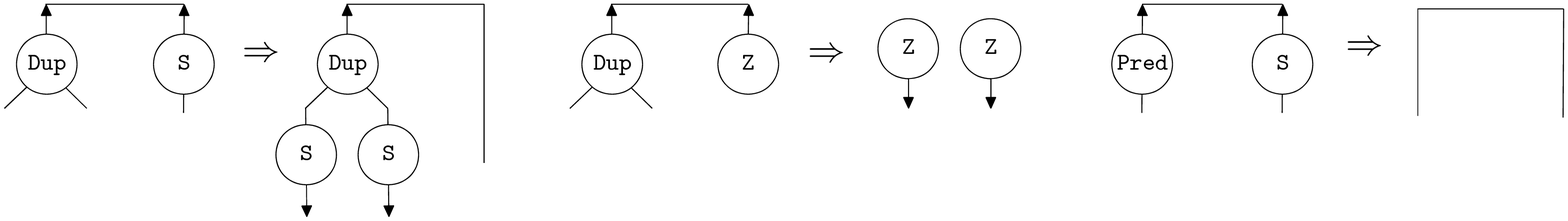}
\end{center}
The following is an example of rewriting:
\begin{center}
\includegraphics[scale=\smallscale,keepaspectratio,clip]{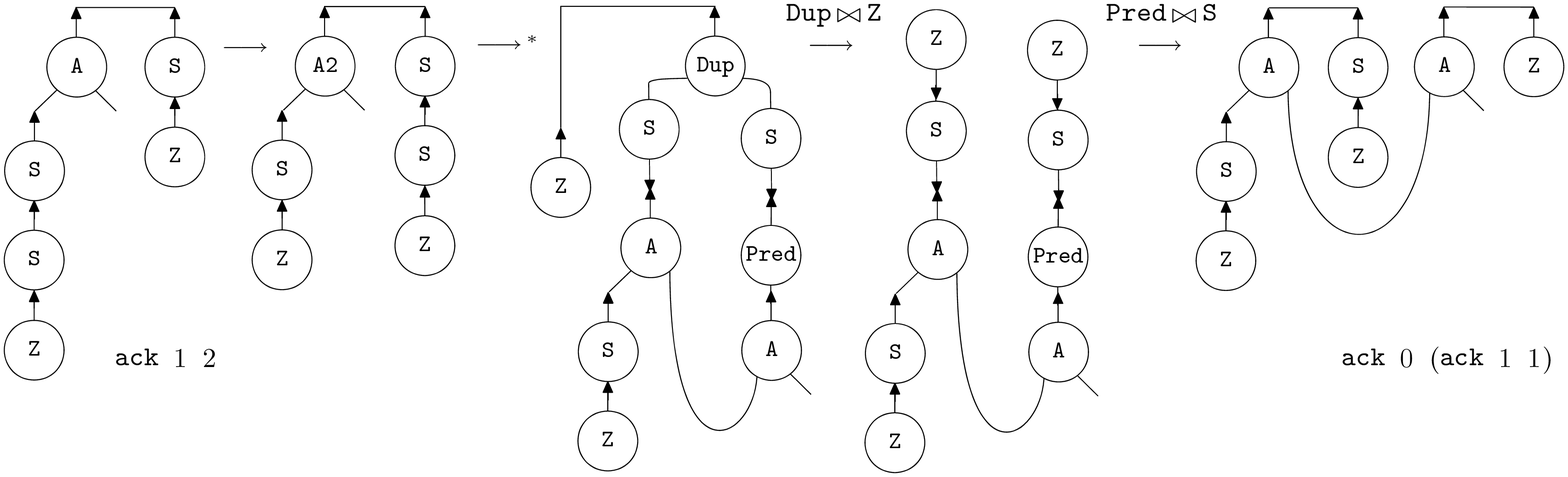}
\end{center}

\paragraph{Observation for in-place running.}
Interaction nets are quite unique as a programming paradigm because we
are basically writing programs using the internal data-structure.  We
characterise three kinds of rewriting rule:

\begin{itemize}

\item Case 1: there are two nodes in the right-hand side (RHS).  The
  two nodes of the active pair can be reused.  Thus, no matter which
  way we evaluate, the algorithm for these rules
  can run in constant space. The rules
  $\add\inter\Succ$, $\sym{A}\inter\Succ$ and
  $\sym{Dup}\inter\Zero$ are classified in this case.

\item Case 2: there are less than two nodes in the RHS.  The active
  pair nodes can be reused as nodes that occur in the RHS, so in terms of
  the memory space, it can run in constant space as well.  For
  instance, the rules $\add\inter\Zero$, $\sym{A}\inter\Zero$ and
  $\sym{Pred}\inter\Succ$ are classified in this case.

\item Case 3: there are more than two nodes in the RHS.  Here, active
  pair nodes can be reused as nodes that occur in the RHS, but
  additional memory space is required for other nodes.  We
  divide this into two very different categories:

  \begin{enumerate}
  \item An active pair creates another active pair that is Case 2
    above. These two reductions together make the algorithm in-place.
    For instance, in the last two-step reductions of $\texttt{ack 1
      2}$ to $\texttt{ack 0 (ack 1 1)}$ in the example, we can save
    memory space for two nodes when we take $\texttt{Pred}\inter\Succ$
    and $\texttt{Dup}\inter\Zero$ in this order, in comparison with
    the order $\texttt{Dup}\inter\Zero$ and
    $\texttt{Pred}\inter\Succ$.
    
  \item For instance, the rules $\sym{A2}\inter\Zero$,
    $\sym{A2}\inter\Succ$ and $\sym{Dup}\inter\Succ$ are 
    classified in this case.  These rules
    are not in-place, but the total cost can be reduced by choosing
    the two reused nodes well.  For instance, in the rule
    $\sym{A2}\inter\Succ$, it is better to reuse the $\sym{A2}$ in the
    LHS as the right side $\sym{A}$ (not the left side) because the
    information of the right auxiliary port (denoted as ``$r$'')
    can be reused. Thus we try to reuse the memory in the best way
    possible.

  \end{enumerate}

\end{itemize}

Insertion sort is a well-known in-place algorithm.  The first three
interaction rules below encode insertion of an item into a sorted
list, and the final two rules encode the insertion sort algorithm.

\begin{center}
\begin{mnet}{190}{20}
\putbox{10}{0}{20}{20}{I$(x)$}
\putbox{50}{0}{20}{20}{\nil}
\putHline{0}{10}{10}
\putRvector{30}{10}{10}
\putLvector{50}{10}{10}
\puttext{95}{10}{$\Lra$}
\putbox{120}{00}{20}{20}{$x$}
\putbox{160}{00}{20}{20}{\nil}
\putLvector{120}{10}{10}
\putLvector{160}{10}{10}
\putHline{140}{10}{10}
\end{mnet}

\begin{mnet}{190}{60}
\putbox{10}{30}{20}{20}{I$(x)$}
\putbox{50}{30}{20}{20}{$y$}
\putHline{0}{40}{10}
\putRvector{30}{40}{10}
\putLvector{50}{40}{10}
\putHline{70}{40}{10}
\puttext{95}{40}{$\stackrel{x\le y}{\Lra}$}
\putbox{120}{30}{20}{20}{$x$}
\putbox{160}{30}{20}{20}{$y$}
\putLvector{120}{40}{10}
\putHline{140}{40}{10}
\putLvector{160}{40}{10}
\putHline{180}{40}{10}
\putbox{10}{0}{20}{20}{I$(x)$}
\putbox{50}{0}{20}{20}{$y$}
\putHline{0}{10}{10}
\putRvector{30}{10}{10}
\putLvector{50}{10}{10}
\putHline{70}{10}{10}
\puttext{95}{10}{$\stackrel{x>y}{\Lra}$}
\putbox{120}{0}{20}{20}{$y$}
\putbox{160}{0}{20}{20}{I$(x)$}
\putLvector{120}{10}{10}
\putHline{140}{10}{20}
\putRvector{180}{10}{10}
\end{mnet}
\end{center}

\begin{net}{190}{20}
\putbox{10}{0}{20}{20}{IS}
\putbox{50}{0}{20}{20}{\nil}
\putHline{0}{10}{10}
\putRvector{30}{10}{10}
\putLvector{50}{10}{10}
\puttext{95}{10}{$\Lra$}
\putbox{120}{00}{20}{20}{\nil}
\putLvector{120}{10}{10}
\end{net}

\begin{net}{190}{20}
\putbox{10}{0}{20}{20}{IS}
\putbox{50}{0}{20}{20}{$x$}
\putHline{0}{10}{10}
\putRvector{30}{10}{10}
\putLvector{50}{10}{10}
\puttext{95}{10}{$\Lra$}
\putbox{120}{00}{20}{20}{I$(x)$}
\putbox{160}{00}{20}{20}{IS}
\putRvector{140}{10}{10}
\putRvector{180}{10}{10}
\putHline{110}{10}{10}
\putHline{70}{10}{10}
\putHline{150}{10}{10}
\end{net}

These five rules encode the whole program---there is nothing else. Our
real point however, is that in this case a trace of the execution (an
animation of this algorithm) is showing no more and no less than what
is needed to explain this algorithm. It is in-place because to begin
with we need to put an additional $\mathrm{IS}$ node, and the final rule for
$\mathrm{IS}$ erases this. We invite the reader to trace the following example
net:
\begin{net}{190}{20}
\putbox{10}{0}{20}{20}{IS}
\putHline{0}{10}{10}
\putRvector{30}{10}{10}
\putbox{50}{0}{20}{20}{2}
\putLvector{50}{10}{10}
\putbox{80}{0}{20}{20}{4}
\putLvector{80}{10}{10}
\putbox{110}{0}{20}{20}{1}
\putLvector{110}{10}{10}
\putbox{140}{0}{20}{20}{3}
\putLvector{140}{10}{10}
\putbox{170}{0}{20}{20}{\nil}
\putLvector{170}{10}{10}
\end{net}

This example is the full system of interaction nets for the insertion
sort algorithm, and it runs in-place. There are other examples, for
instance reversing a list. In this case, we need to start with adding
a \rev\ node, and the final rule deletes it. The following two rules
implement reversing any list:

\begin{center}
\begin{mnet}{130}{40}
    \putbox{10}{00}{20}{30}{\rev}
    \putHline{0}{15}{10}
    \putHline{30}{10}{10}
    \putbox{50}{10}{20}{20}{\nil}
    \putLvector{50}{20}{10}
    \putRvector{30}{20}{10}
    \puttext{90}{20}{$\Lra$}
    \putHline{110}{20}{20}
\end{mnet}
\qquad\qquad
\begin{mnet}{190}{40}
  \putbox{10}{0}{20}{30}{\rev}
  \putbox{130}{0}{20}{30}{\rev}
  \putHline{0}{15}{10}
  \putHline{30}{10}{10}
  \putbox{50}{10}{20}{20}{$h$}
  \putbox{170}{0}{20}{20}{$h$}
  \putHline{120}{15}{10}
  \putHline{150}{10}{10}
  \putHline{190}{10}{10}
  \putLvector{170}{10}{10}
  \putLvector{50}{20}{10}
  \putRvector{30}{20}{10}
  \putHline{190}{10}{10}
  \puttext{100}{20}{$\Lra$}
  \putRvector{150}{20}{10}
  \putHline{70}{20}{10}
\end{mnet}
\end{center}

The starting configuration is shown in the following example, which will reverse the list in-place with five interactions:
\begin{net}{190}{60}
\putbox{10}{0}{20}{30}{\rev}
\putHline{0}{15}{10}
\putbox{50}{30}{20}{20}{1}
\putbox{80}{30}{20}{20}{2}
\putLvector{80}{40}{10}
\putbox{110}{30}{20}{20}{3}
\putLvector{110}{40}{10}
\putbox{140}{30}{20}{20}{4}
\putLvector{140}{40}{10}
\putbox{170}{30}{20}{20}{\nil}
\putLvector{170}{40}{10}
\putHline{30}{10}{10}
\putbox{50}{0}{20}{20}{\nil}
\putLvector{50}{10}{10}
\qbezier(30,20)(40,20)(40,30)
\qbezier(40,30)(40,40)(50,40)
\putUvector{40}{30}{1}
\putDvector{40}{30}{1}
\end{net}

\section{Annotating Rules}\label{sec:annotation}

In the previous section we saw that there are interaction systems that
can run in-place. To ensure that this is actually achieved at the
implementation level, we need to make sure that for these examples the
nodes in the rule are reused when building the new net. This idea can
be also used even when there are more than two nodes in the right-hand
side of the rule, and in this case there is a choice of how to reuse
the nodes. We next introduce an annotation to show which nodes in
the RHS of the rule are reused.  This helps the compiler to analyse
information so that it can improve on the in-place execution of parts
of the program.

\paragraph{Annotation: $\sym{*L}$ and $\sym{*R}$.}
We introduce $\sym{*L}$ and $\sym{*R}$ to denote where the left-hand
side and the right-hand side nodes in the LHS of a rule are used for
in-place computation, respectively.  For instance, the rule
$\add\inter\Succ$ is written as follows:
\begin{center}
\includegraphics[scale=\normalscale,keepaspectratio,clip]{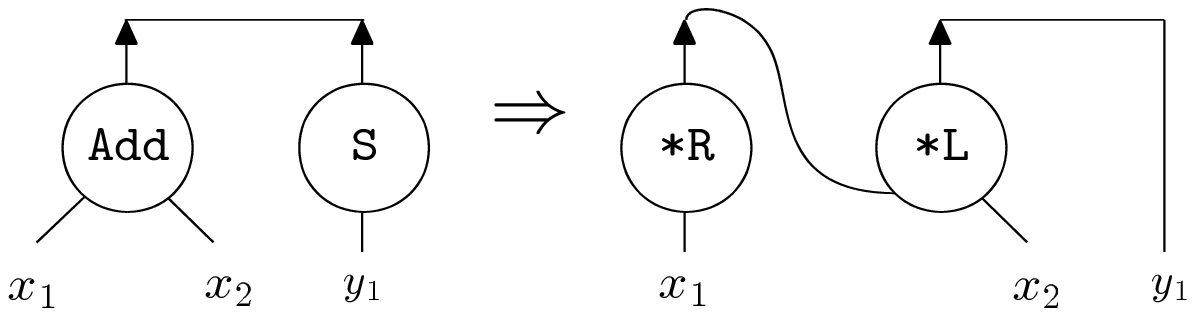}
\end{center}

The advantage is that the compiler is easily able to know, traversing
the net, not only where the active pair nodes are used, but also which
information about the connections should be preserved.  For instance,
in the above example, the information denoted as ``$x_1$'' and
``$y_1$'' in the LHS of the rule should be preserved, say as
``\verb|_|$x_1$'' and ``\verb|_|$y_1$'', because these are overwritten
in the RHS of the rule, and the ``$x_1$'' and ``$y_1$'' in the RHS
should be replaced by the ``\verb|_|$x_1$'' and ``\verb|_|$y_1$''.

\paragraph{Annotation for changing of the node name.}
To change the node name, we introduce a name cast, such as the type cast
of the C programming language, with the $\sym{*L}$ and $\sym{*R}$.
For instance, the rule $\sym{A2}\inter\Succ$ is written as follows:
\begin{center}
\includegraphics[scale=\normalscale,keepaspectratio,clip]{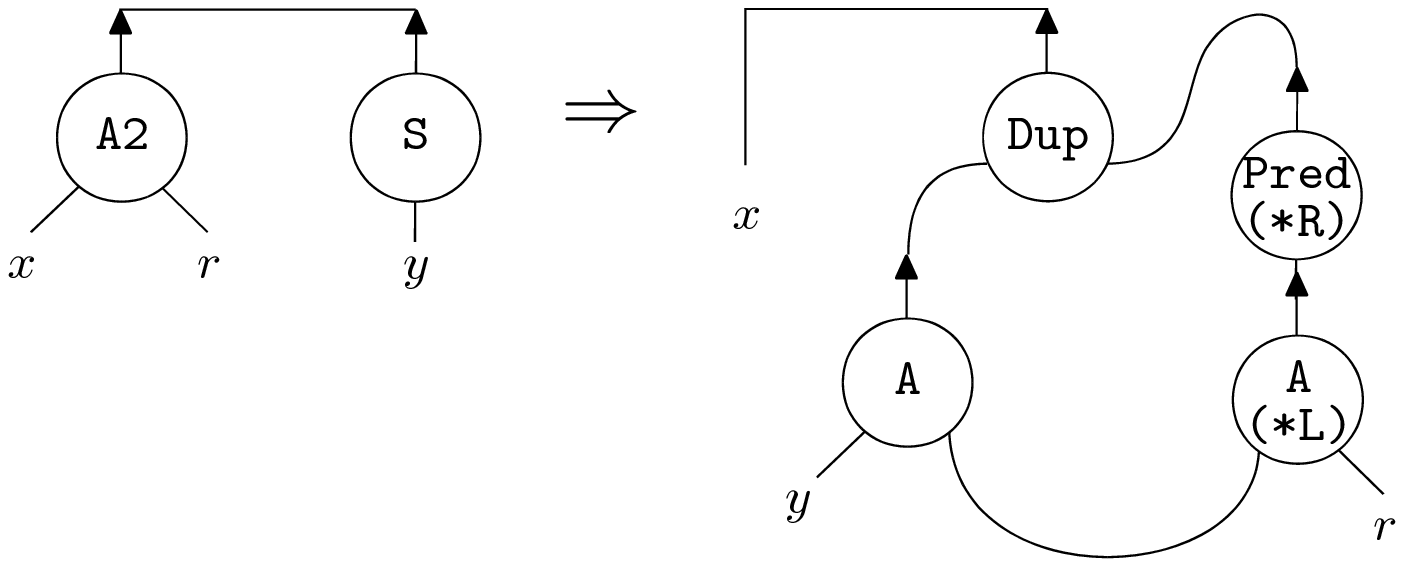}
\end{center}

In this example, the compiler can also know that the information denoted
as ``$x$'' and ``$y$'' in the LHS of the rule should be preserved, by
checking the connection of $\sym{*L}$ and $\sym{*R}$.

\paragraph{Advantage of using a fixed-size node representation for nodes.}
We represent nodes as a fixed-size node, thus fixed-size auxiliary
ports. For this we need to use more space than necessary, but we can
manage and reuse nodes in a simpler way~\cite{PeytonJonesSL:impfpl}.
Here, we assume that auxiliary ports are assigned by the order from
the left-hand side to the right-hand side, the rule
$\sym{A}\inter\Zero$ is written as follows:
\begin{center}
\includegraphics[scale=\normalscale,keepaspectratio,clip]{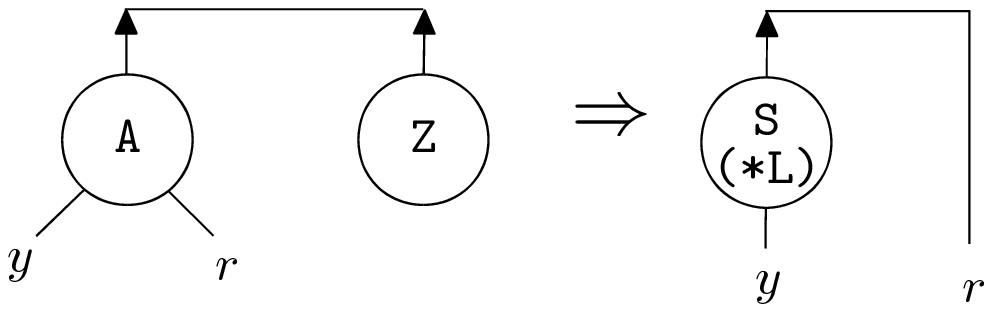}
\end{center}
So we can reuse not only the $\sym{*L}$ node, but also the pointer
information (denoted as ``$y$'').

\section{Deriving Annotations}\label{sec:deriving}

In the previous section we showed how rules can be annotated with
information about the reuse of nodes.  Here, in the case of using a
fixed-size node representation, we define a function to calculate how
a given term is similar to others.  We first introduce some
notation. For strings we write double quotes ($\delimStrS$ and
$\delimStrE$).  We use the notation $\{x\}$ in a string as the result
of replacing the occurrence $\{x\}$ with its actual value.  For
instance, if $x=\Str{\sym{abc}}$ and $y=89$ then
$\Str{\sym{1}\{x\}\sym{2}\{y\}} = \Str{\sym{1abc289}}$.  We use $+$ as
an infix binary operation to concatenate strings.
We also use a symbol $\Empty$ to show the empty sequence.
In order to show where in a term is in a sequence of equations, we use
the following term path notation: $\mathit{nth} \ (\sym{L}\mid\sym{R})
\ : \mathit{arg_1} \ \mathit{arg_2} \ldots$. For instance, the term
$t$ in $x \EQ y, \alpha(\beta(s,t),z) \EQ w$ is denoted as
\texttt{2L:12} because the $t$ occurs in $\alpha(\beta(s,t),z)$, which
is the left-hand side term of the second equation, occurs in
$\beta(s,t)$, which is the first argument of the term, and in the
second argument of the term $\beta(s,t)$.

Using this, we can now give an important definition for this paper:

\begin{definition}[Node matching]
The function $\Func{Match}$ below takes a term and a sequence of
equations, and returns a list of a pair $\sym{(}\mathit{score}\sym{,
}\mathit{\ term\ path}\sym{)}$ where the score contains the number of
the matched agent and matched arguments. We use Standard ML notation
for lists, thus, in the following the operator ``@'' is list
concatenation.

\medskip

$
\begin{array}{lcl}
\Func{Match}(t,\, (e_1,\ldots,e_n)) &=& \Func{Match}_e(t,\, e_1,\, 1) @ \cdots  @ \mathrm{Match}_e(t,\, e_n,\, n) \\
\end{array}
$
\medskip

$
\begin{array}{lcl}
\Func{Match}_e(t,\, s=u,\, \mathit{pos}) &=& \Func{Match}_t(t,\, s,\, \Str{\{\mathit{pos}\}\sym{L}:}) @  \Func{Match}_t(t,\, u,\, \Str{\{\mathit{pos}\}\sym{R}:})
\end{array}
$
\medskip

$\begin{array}{llcl}
&\Func{Match}_t(\alpha(\myvec{x}),\, x,\, \mathit{tpath}) &=& 
[((0,0), \mathit{tpath})]\\
\mid& \Func{Match}_t(\alpha(\myvec{x}),\, \beta(t_1,\ldots,t_n),\, 
\mathit{tpath}) &=& 
[((\mathit{agentPts},\, \mathit{namePts}),\, \mathit{tpath})]\\
&&&@ \Func{Match}_t(\alpha(\myvec{x}),\, t_1,\, \Str{\{\mathit{tpath}\}\sym{1}}) @ \cdots \\
&&& @ \Func{Match}_t(\alpha(\myvec{x}),\, t_n,\, \Str{\{\mathit{tpath}\}\sym{n}} )\\
\multicolumn{4}{r}{\mathsf{where} \quad \mathit{agentPts} \,=\, \Func{if} \ \alpha=\beta \ \Func{then} \ 1 \ \Func{else} \ 0} \\
\multicolumn{4}{r}{\mathit{namePts} \,=\, \Func{Match}_{ns}(\myvec{x},\, (t_1,\ldots,t_n))}\\
\end{array}
$
\medskip

$\begin{array}{llcl}
&\Func{Match}_{ns}(\Empty,\, \myvec{t}) &=& 0 \\
\mid &\Func{Match}_{ns}(\myvec{x},\, \Empty) &=& 0 \\
\mid &\Func{Match}_{ns}((x,\myvec{x}),\, (t,\myvec{t})) &=&  
(\Func{if} \ x=t \ \Func{then} \ 1 \ \Func{else} \ 0) +
\Func{Match}_{ns}(\myvec{x},\, \myvec{t})\\
\end{array}
$
\end{definition}

Thus \texttt{Match} will return a list of the matching metrics for
each node, and give the location in the net for each. We can then
easily extract the best one from this information.  The following
examples illustrate how this information is calculated.

\begin{example}
The rule $\sym{Add}\inter\sym{S}$ given previously is written textually as
\[
\sym{Add}(x_1,x_2) \EQ \sym{S}(y_1) \ito 
\sym{Add}(\sym{S}(x_1,x_2)) \EQ y_1
\]
First we take the left hand side term $\sym{Add}(x_1,x_2)$ of the active pair.
\[\Func{Match}(\sym{Add}(x_1,x_2), \ \sym{Add}(\sym{S}(x_1),x_2) \EQ y_1)\]
returns the following list:
\begin{alltt}
[((1,1),1L:), ((0,1),1L:1), ((0,0),1L:11), ((0,0),1L:2), ((0,0),1R:)]
\end{alltt}

\noindent
The first element of this result $\sym{((1,1),1L:)}$ shows that
the highest score $\sym{(1,1)}$ is obtained when we annotate
the term $\sym{1L:}$,
which is $\sym{Add}(\sym{S}(x_1),x_2)$, such as $\sym{(*L)}(\sym{S}(x_1),x_2)$.
Next we take the right hand side term $\sym{S}(y_1)$. The following is the result of the function $\Func{Match}$ for $\sym{S}(y_1)$:
\begin{alltt}
[((0,0),1L:), ((1,0),1L:1), ((0,0),1L:11), ((0,0),1L:2), ((0,0),1R:)]
\end{alltt}
This result shows that there is one agent term $\sym{1L:1}$ (thus
$\sym{S}(x_1)$) that has the same id with $\sym{S}(y_1)$, and no agent
terms that have the same occurrence of the argument.  Therefore,
$\sym{S}(x_1)$ should be annotated as $\sym{(*R)}(x_1)$.  These
annotations correspond to the first graph in
Section~\ref{sec:annotation}.
\end{example}

\begin{example}
The rule $\sym{A2}\inter\sym{S}$ given previously is written textually
as
\[
\sym{A2}(x,r) \EQ \sym{S}(y) \ito x \EQ
\sym{Dup}(\sym{A}(y,w),\sym{Pred}(\sym{A}(w,r)))
\]
The result of $\Func{Match}(\sym{A2}(x,r), \ x \EQ
\sym{Dup}(\sym{A}(y,w),\sym{Pred}(\sym{A}(w,r))))$ is as follows:
\begin{alltt}
 [((0,0),1L:), ((0,0),1R:), ((0,0),1R:1), ((0,0),1R:11), ((0,0),1R:12),
  ((0,0),1R:2), ((0,1),1R:21)), ((0,0),1R:211), ((0,0),1R:212)].
\end{alltt}

\noindent
This shows that the term $\sym{1R:21}$, which is $\sym{A}(w,r)$,
should be annotated by $\sym{(*L)}$ because it has the highest score
$\sym{(0,1)}$.

In the case of $\Succ(y)$, the result of the $\Func{Match}$ is as follows:
\begin{alltt}
 [((0,0),1L:), ((0,0),1R:), ((0,1),1R:1), ((0,0),1R:11), ((0,0),1R:12),
  ((0,0),1R:2), ((0,0),1R:21)), ((0,0),1R:211), ((0,0),1R:212)].
\end{alltt}
  
\noindent
Thus, taking account of using a fixed-size node representation, we
find that we should annotate the term $\sym{1R:1}$, which is
$\sym{A}(y,w)$, by $\sym{(*R)}$, though we annotated $\sym{Pred}$ in
the graph in Section~\ref{sec:annotation}.  Of course, the evaluation
of the score depends on the implementation method.  However, in these
cases, $\sym{(0,1)}$ must be highest because the others are
$\sym{(0,0)}$.
\end{example}

\section{Discussion}\label{sec:discussion}

The algorithm given above can be put to use in the compilation of
rules. There is some choice for the compiler if several nodes get the
same score. We briefly summarise here some details about the low-level
language, and a description of how data-structures are used. Our
contribution here is to extend those ideas with reuse. In the longer
version of this paper we will give the algorithm for compilation in
detail and focus on the data-structures that are used. An important
result that we get about compilation is that we do the least amount of
work in implementing the interaction rules.

The concrete representation of an interaction net can be summarised by
the following diagram, where $\Gamma$ represents a net, $\sym{EQ}$ a
stack of equations, and $\sym{I}$ an interface of the net:
\begin{center}
\includegraphics[width=6.2cm,keepaspectratio,clip]{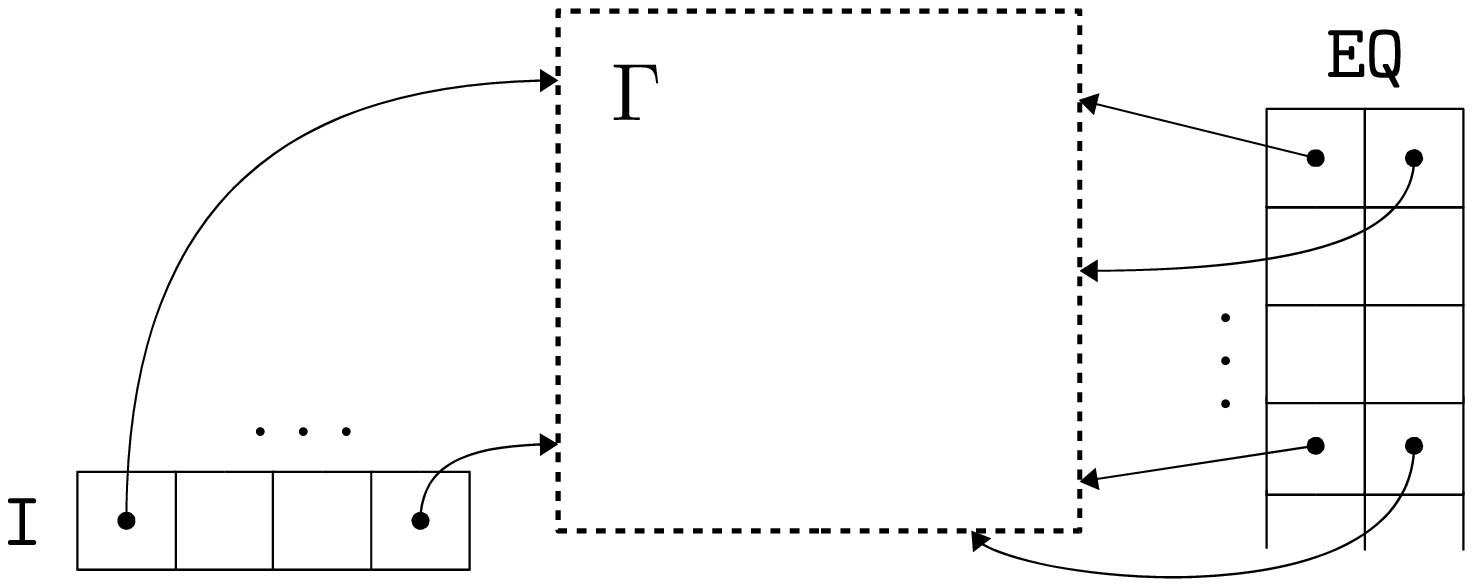}
\end{center}

The following is the representation of the net
$\sym{Add}(\sym{Z},r)=\sym{S}(w), \,
\sym{Add}(\sym{Z},w)=\sym{S}(\sym{Z})$, where $\sym{N}$ are nodes
representing variables:

\begin{center}
\includegraphics[scale=\tinyscale,keepaspectratio,clip]{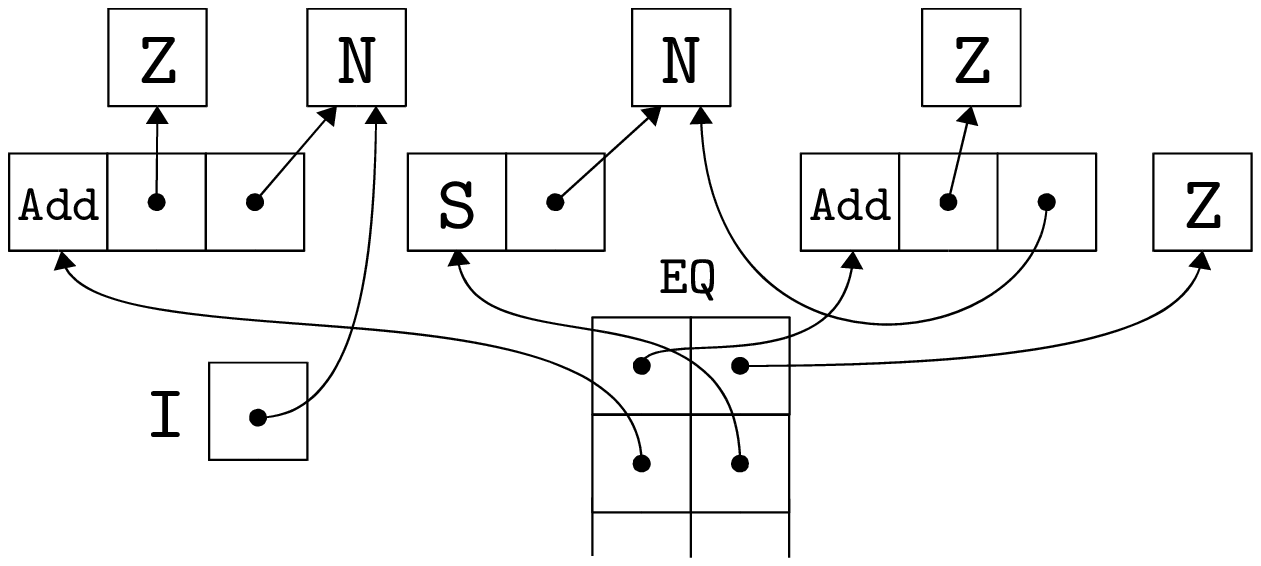}
\end{center}

The interface of the net is where we collect all the free edges of a
net. We have a set of instructions that can manipulate this
data-structure, and the compilation of a rewrite rule needs to
generate code to manipulate this structure. What we have achieved in
this paper is an annotation of a rule so that we can generate code to
manipulate data-structures of the kind shown above with the least
amount of work. Specifically, we limit the allocation of memory on the
heap, and also avoid unnecessary garbage collection. We also find the
best way to build the RHS of the rule, by re-using the memory cells
and the existing pointers.

Returning to the ideas presented in the introduction, we have obtained
the following:

\begin{itemize}

\item Case 1: if there are two nodes in the RHS of the rule, then the
  algorithm for executing these rules will run in-place, and moreover
  we will perform the fewest updates of pointers to build the
  RHS. This is because the algorithm above will find the best way of
  reusing both nodes and connections.
  
\item Cases 2 and 3: Even when the algorithm is not in-place because
  we have more than two nodes or fewer than two nodes in the RHS, the
  ideas of this paper still gives the optimal implementation of the
  rule. We note however that if we change the data-structure, there
  may be other solutions to this problem, so the result we obtain is
  with respect to the chosen data-structure.

\end{itemize}

One aspect of in-place interaction nets reduction that we have not
examined in the current paper is pairing up interaction rules by using
a specific reduction strategy. For example, if an interaction system
has a rule with one node in the RHS, and another rule with three, then
we might be able to schedule these two rules to be performed together,
this maintaining the algorithm in-place. There are a number of ways we
can do this at the implementation level, and we will include details
of this aspect in the long version of the paper.

\section{Conclusion}\label{sec:conc}

We have introduced a notion to rewrite graphs, specifically
interaction nets, in-place. We have identified a number of in-place
algorithms and this lead us to an annotation to facilitate the
implementation of in-place rewriting by re-using nodes of active
pairs. The main feature of this annotation is that is will give the
best possible reuse, so will allow the least amount of work to be
done when rewriting the graph.  We are currently working to
incorporate the ideas of this paper into an implementation, using the
model previously developed in~\cite{DBLP:journals/corr/HassanMS15}. We
hope to present details of that, and benchmark results at a future time.

\bibliographystyle{eptcs} 
\bibliography{bibfile}

\end{document}